\documentclass[11pt]{iopart}
\usepackage{iopams}  
\usepackage{amssymb}
\usepackage{graphicx}
\usepackage{graphics}
\usepackage{epstopdf}
\begin{document}

\title{Laser stimulated thermal conductivity in chiral carbon nanotube}

\author{A. Twum$^1$, S. Y. Mensah$^1$, N. G. Mensah$^2$, K. A. Dompreh$^1$, S. S. Abukari$^1$ and M. Rabiu$^3$}
\address{$^1$ Department of Physics, Laser and Fiber Optics Center, University of Cape Coast, Cape Coast, Ghana.}
\address{$^2$ Department of Mathematics, University of Cape Coast, Cape Coast, Ghana.}
\address{$^3$ Department of Applied Physics, Faculty of Applied Sciences, University for Development Studies, Navrongo Campus, Ghana.}
\ead{+233 042 33837, profsymensah@yahoo.co.uk}

\begin{abstract}
An investigation of laser stimulated thermal conductivity in chiral CNT is presented. The thermal conductivity of a chiral CNT is calculated using a tractable analytical approach. This is done by solving the Boltzmann transport equation with energy dispersion relation obtained in the tight binding approximation. The electron thermal conductivity along the circumferential $\chi_{c}$ and axial $\chi_{z}$ are obtained. The results obtained are numerically analyzed and both $\chi_{c}$ and $\chi_{z}$ are found to oscillate in the presence of laser radiations. We have also noted that the laser source caused a drastic reduction in the both $\chi_{c}$ and $\chi_{z}$ values.
\end{abstract}


\maketitle 

\section{Introduction}
Carbon-based materials (diamond and in-plane graphite) display the highest measured thermal conductivity of any known material at moderate temperatures \cite{1}. The discovery of carbon nanotubes in 1991 \cite{2} has led to speculation that this new material could have a thermal conductivity greater than that of diamond and graphite [3]. Carbon nanotube has found a lot of application in electronic and mechanical devices. It is, therefore, not surprising that the material has received a lot of attention over the past decade [4-11].

 The thermal conductivity of materials in general is partitioned into charge carriers (i.e., electron or hole) component $\chi _{\rm e}$ which depends on the electronic band structure, electron scattering and electron-phonon interaction, and lattice component $\chi _{\rm L }$ which depends mainly on phonon and phonon scattering. In dielectrics, $\chi_ L \gg  \chi_e$ while in metals the reverse is the case. In semiconductors, the value of the thermal conductivity $\chi $ is strongly dependent on the composition of the semiconductor, and the value of $\chi_L$ is generally greater than the value of $\chi_e$.

So far, most publications on the thermal conductivity of carbon nanotubes have paid attention to only the lattice thermal conductivity and completely neglected electron thermal conductivity. Hone et al. \cite{3} found that the conductivity of carbon nanotubes was temperature dependent, and was almost a linear relationship. They suggested that the conductivity decreases smoothly with temperature, and displays linear temperature dependence below 30 K. However, Berber et al. \cite{12} suggested that the graph of the temperature dependence of thermal conductivity looked less linear and that it shows a positive slope from low temperatures up to 100 K where it peaks around 37000 W/mK. Then, the thermal conductivity drops dramatically down to around 3000 W/mK when the temperature approaches 400 K. Similar relationship has been found by Mensah et al. \cite{13} for electron thermal conductivity $\chi_e$.  

Mensah et al. \cite{14} have also studied the electron thermal conductivity of carbon nanotubes. They observed that the temperature dependence of $\chi_e$ in carbon nanotubes is similar to that obtained by Berber et al. and that $\chi_e$ peaks at unusually high values. They further observed the dependence of $\chi_e$ on the geometric chiral angle $\theta $, temperature T, the real overlapping integrals for jumps along the tubular axis $\Delta_{z} $  and the base helix $\Delta_{s}$. Interestingly, they again noted that varying these parameters could give rise to unusual high electron thermal conductivity whose peak values shift towards higher temperatures. For example, at $\Delta_{ z} = 0.02$ eV and $\Delta_{s} = 0.015$ eV. The peak value of $\chi_e$ occurs at 104K and is about 41000 W/mK which compares well with that reported for a 99.9\% isotropically enriched 12C diamond crystal. In this work, we will use the approach in \cite{15} to investigate theoretically the laser stimulated thermal conductivity in chiral CNTs. In this paper we consider the effect of  laser on the thermal conductivity of  chiral carbon nanotube. We observed that the laser has drastic effect on the electron thermal conductivity. It drastically reduced the thermal conductivity i.e. about 10 times. It also causes $\chi $  to oscillate with the amplitude of the laser source.

The paper is organized as follows: section one deal with the introduction; in section two, we establish the theory and solutions; results obtained will be discussed in section three and finally we draw our conclusions

\section{Theory}

The thermal current density q and electron thermal conductivity $\chi$ of a chiral SWNT are calculated as functions of the geometric chiral angle $\theta_{h}$, temperature T, the real overlapping integrals for jumps along the nanotube axis $\Delta_{z}$ and along the base helix $\Delta_{s}$. The calculation is done using the approach in reference [15] together with the phenomenological model of a SWNT developed in references \cite{16} and \cite{17}. This model yields physically interpretable results and gives correct qualitative descriptions of various electronic processes, which are corroborated by the first-principle numerical simulations of Miyamoto {\it et. al.} \cite{18}.

Following the approach of \cite{19}, we consider a SWNT under a temperature gradient $\nabla T$ and placed in an electric field applied along the nanotube axis. Employing the Boltzmann kinetic equation 
\begin{equation} \label{EQ_1} 
	\frac{\partial f(r, p, t)}{\partial t} +v(p)\frac{\partial f(r, p, t)}{\partial r} + eE(t)\frac{\partial f(r, p, t}{\partial p} = \frac{\partial (r, p, t) - f_{0}(p)}{\tau }  
\end{equation} 
where $f(r, p, t)$ is the distribution function, $f_0(p)$ is the equilibrium distribution function, $v(p)$ is the electron velocity, $E(t)= E_0 + E_1 cos(\omega t)$ is the magnitude of the electric field, with $E_0$ being constant electric field and $E_1cos(\omega t)$ being monochromatic laser source, $r$ is the electron position, $p$ is the electron dynamical momentum, $t$ is time elapsed,$\tau $ is the electron relaxation time and $e$ is the electron charge and taken the collision integral in the $\tau$ approximation and further assumed constant, the exact solution of \eref{EQ_1} is solved using perturbation approach where the second term is treated as the perturbation. In the linear approximation of $\nabla T$ and $\nabla\mu$, the solution to the Boltzmann kinetic equation is 
\begin{eqnarray} 
\fl	f(p, t) = \tau ^{-1} \int _{0}^{\infty }\exp  \left(-\frac{t}{\tau } \right)f_{0} \left(p-e\int _{t-t^{'} }^{t}\left[E_{0} +E_{1} \cos \omega t^{''} \right]dt^{''}  \right)dt + \cdots \nonumber\\
		 \int _{0}^{\infty }\exp \left(-\frac{t}{\tau } \right) dt\left\{\left[\varepsilon \left(p-e\int _{t-t^{'} }^{t}\left[E_{0} +E_{1} \cos wt^{''} \right]dt^{''}  \right)-\mu \right]\frac{\nabla T}{T} +\nabla \mu \right\} \nonumber\\
		 \times v\left(p-e\int _{t-t^{'} }^{t}\left[E_{0} +E_{1} \cos \omega t^{''} \right]dt^{''}  \right)\nonumber\\
		 \times \frac{\partial f_{0} }{\partial \varepsilon } \left(p-e\int _{t-t^{'} }^{t}\left[E_{0} +E_{1} \cos wt^{''} \right]dt^{''}  \right) \label{EQ_2} 
\end{eqnarray} 
$\varepsilon (p)$ is the tight-binding energy of the electron, and $\mu$ is the chemical potential. The thermal current density q is defined as
\begin{equation} 
	q = \sum_{p}\left[\varepsilon (p) - \mu \right]v(p)f(p).\label{EQ_3} 
\end{equation} 
Substituting \eref{EQ_2} into \eref{EQ_3} we have
\begin{eqnarray} 
\fl	q = \tau^{-1} \int _{0}^{\infty }\exp(-\frac{t}{\tau } ) dt\sum _{p}[\varepsilon \left(p\right)-\mu]v(p) f_{0} \left(p-e\int _{t-t'}^{t}\left[E_{0} +E_{1} \cos \omega t''\right] dt''\right)\nonumber\\
	    +\int_0^{\infty }\exp \left(-\frac{t}{\tau } \right) dt\sum_{p}\left[\varepsilon \left(p\right)-\mu \right]v\left(p\right)\nonumber\\
	    \times \left\{\left[\varepsilon \left(p-e\int _{t-t'}^{t}\left[E_{0} +E_{1} \cos \omega t''\right] dt''-\mu \right)\right]\frac{\nabla T}{T} +\nabla \mu \right\}\nonumber\\
	     \times v\left(p-e\int _{t-t'}^{t}\left[E_{0} + E_{1} \cos \omega t''\right] dt''\right)\frac{\partial f_{0} }{\partial \varepsilon } \left(p-e\int _{t-t'}^{t}\left[E_{0} + E_{1} \cos \omega t''\right] dt''\right).\label{EQ_4} 
\end{eqnarray} 
Making the transformation 
\[
	p-e\int _{}^{}\left[E_{0} +E_{1} \cos wt''\right]dt''\to p, 
\] 
we obtain for the thermal current density
\begin{eqnarray} 
\fl	q = \tau^{-1} \int _{0}^{\infty }\exp \left(-\frac{t}{\tau } \right) dt\sum _{p}\left[\varepsilon \left(p-e\int _{t-t'}^{t}\left[E_{0} + E_{1} \cos wt''\right] dt''\right)-\mu \right] \nonumber\\
	\times v\left(p-e\int _{t-t'}^{t}\left[E_{0} +E_{1} \cos wt''\right] dt''\right)f_{0} \left(p\right)\ldots\nonumber\\
	+\int_{0}^{\infty }\exp \left(-\frac{t}{\tau } \right) dt\sum _{p}\left[\varepsilon \left(p-e\int _{t-t'}^{t}\left[E_{0} + E_{1} \cos \omega t''\right] dt''\right)-\mu \right] \nonumber\\
	\times \left\{\left[\varepsilon \left(p\right)-\mu \right]\frac{\nabla T}{T} +\nabla \mu \right\} \left\{v\left(p\right)\frac{\partial f_{0} \left(p\right)}{\partial \varepsilon } \right\}\nonumber\\
	\times v\left(p-e\int _{t-t'}^{t}\left[E_{0} +E_{1} \cos \omega t''\right] dt''\right)  \label{EQ_5} 
\end{eqnarray} 

Using the phenomenological model in \cite{16, 17, 20}, a SWNT is considered as an infinitely long periodic chain of carbon atoms wrapped along a base helix and the thermal current density is written in the form 
\begin{equation} 
	q=S' u_{s} +Z'u_{z}  \label{EQ_6} 
\end{equation} 
where $S'$ and $Z'$ are respectively components of the thermal current density along the base helix and along the nanotube axis. The motion of electrons in the SWNT is resolved along the nanotube axis in the direction of the unit vector $u_{z}$ and a unit vector $u_{s}$ tangential to the base helix. $u_{c}$ is defined as the unit vector tangential to the circumference of the nanotube  and $\theta_{h}$ is the geometric chiral angle (GCA). $u_{c}$ is always perpendicular to $u_{z}$, therefore $u_{s}$ can be resolved along $u_{c}$ and $u_{z}$ as 
\begin{equation} 
	u_{s} = u_{c} cos \theta_{h} + u_{z} sin \theta_{h}.\label{EQ_7} 
\end{equation}
Therefore, $j$ can be expressed in terms of $u_{c}$ and $u_{z}$ as
\begin{equation} 
	q = u_{c}(S' cos \theta_{h}) + u_{z} (Z' + S' sin \theta_{h}) \equiv j_{c }u_{c} + j_{z }u_{z}\label{EQ_8}
\end{equation}
which implies that, 
\begin{equation} 
	q_{c} = S' cos \theta_{h}\label{EQ_9}
\end{equation}
\begin{equation} 
	q_{z} = Z' + S' sin \theta_{h}\label{EQ_10}
\end{equation}

The interference between the axial and helical paths connecting a pair of atoms is neglected so that transverse motion quantization is ignored \cite{16, 17}. This approximation best describes doped chiral carbon nanotubes, and is experimentally confirmed in \cite{21}. Thus if in \eref{EQ_5} the transformation
\[
	\sum _{p} \to \frac{2}{ (2\pi \hbar)^{2} } \int_{ -\frac{\pi }{ d_s }}^{ \frac{\pi }{ d_{s} }}dP_{s}\int _{ -\frac{\pi }{ d_z }}^{ \frac{\pi }{ d_z }} dP_{z}  
\]
is made, Z′ and S′ respectively become,
\begin{eqnarray} 
\fl	Z ' =  \frac{2 }{(2\pi \hbar)^{2} } \int _{0}^{\infty }\exp \left(-\frac{t}{\tau } \right) dt \int_{ -\frac{\pi }{ d_s }}^{ \frac{\pi }{ d_{s} }}dP_{s}\int _{ -\frac{\pi }{ d_z }}^{ \frac{\pi }{ d_z }} dP_{z} \left[\varepsilon \left(p-e\int _{t-t'}^{t}\left[E_{0} + E_{z} \cos \omega t''\right] dt''\right)-\mu \right]\nonumber\\
	  \times v_{z} \left(p - e\int _{t-t'}^{t}\left[E_{0} +E_{s} \cos wt''\right] dt''\right)\Big[ \tau^{-1}f_0(p) +  \cdots \nonumber\\
	   \left\{\left[\varepsilon \left(p\right)-\mu \right]\frac{\nabla _{z} T}{T} +\nabla _{z} \mu \right\}\left\{v_{z} \left(p\right)\frac{\partial f_{0} \left(p\right)}{\partial \varepsilon } \right\} \Big]\label{EQ_11}
\end{eqnarray} 
and
\begin{eqnarray} 
\fl	S ' =  \frac{2 }{(2\pi \hbar)^{2} } \int _{0}^{\infty }\exp \left(-\frac{t}{\tau } \right) dt  \int_{ -\frac{\pi }{ d_s }}^{ \frac{\pi }{ d_{s} }}dP_{s}\int _{ -\frac{\pi }{ d_z}}^{ \frac{\pi }{ d_z }} dP_{z}\left[\varepsilon \left(p-e\int _{t-t'}^{t}\left[E_{0} + E_{s} \cos \omega t''\right] dt''\right)-\mu \right]\nonumber\\
	   \times v_{s} \left(p - e\int _{t-t'}^{t}\left[E_{0} +E_{z} \cos wt''\right] dt''\right)\Big[ \tau^{-1}f_0(p) +  \cdots \nonumber\\
	   \left\{\left[\varepsilon \left(p\right)-\mu \right]\frac{\nabla _{s} T}{T} +\nabla _{s} \mu \right\}\left\{v_{s} \left(p\right)\frac{\partial f_{0} \left(p\right)}{\partial \varepsilon } \right\} \Big] \label{EQ_12}
\end{eqnarray} 
where the integrations are carried out over the first Brillouin zone, $\hbar$ is Planck's constant, $v_{s}$, $p_{s}$, $E_{s}$, $\nabla_{s}T$, and $\nabla_{s}\mu$ are the respective components of $v$, $p$, $E$, $\nabla T$ and $\nabla\mu$ along the base helix, and v$_{z}$, $p_{z}$, $E_{z}$, $\nabla_{z}T$, and $\nabla_{z}\mu$ are the respective components along the nanotube axis.

The energy dispersion relation for a chiral nanotube obtained in the tight binding approximation \cite{16} is
\begin{equation}  
	\varepsilon \left(p\right) = \varepsilon_{0} -\Delta _{s} \cos \frac{P_{s} d_{s} }{\hbar } -\Delta _{z} \cos \frac{P_{z} d_{z} }{\hbar}  \label{EQ_13}
\end{equation} 
where $\varepsilon_{0}$ is the energy of an outer-shell electron in an isolated carbon atom, $\Delta_{z}$ and $\Delta_{s}$ are the real overlapping integrals for jumps along the respective coordinates, $p_{s}$ and $p_{z}$ are the components of momentum tangential to the base helix and along the the nanotube axis, respectively. The components $v_{s}$ and $v_{z}$ of the electron velocity $V$ are respectively calculated from the energy dispersion relation \ref{EQ_13} as
\begin{equation} 
	v_{s} \left(p\right)=\frac{\partial \varepsilon \left(p\right)}{\partial P_{s} } =\frac{\Delta _{s} d_{s} }{\hbar } \sin \frac{P_{s} d_{s} }{\hbar } \label{EQ_14}  
\end{equation} 
\begin{eqnarray} 
\fl	v_{s} \left(p-e\int _{t-t^{'} }^{t}\left[E_{0} +E_{1} \cos wt^{''} \right]dt^{''}  \right) = \frac{\Delta _{s} d_{s} }{\hbar } \sin \left(p-e\int _{t-t^{'} }^{t}\left[E_{0} +E_{1} \cos wt^{''} \right]dt^{''}  \right)\nonumber\\
=\frac{\Delta _{s} d_{s} }{\hbar } \left\{\sin \frac{P_{s} d_{s} }{\hbar } \cos \left(p-e\int _{t-t^{'} }^{t}\left[E_{0} +E_{1} \cos wt^{''} \right]dt^{''}  \right)\right\}\nonumber\\
-\cos \frac{P_{s} d_{s} }{\hbar } \sin \left(p-e\int _{t-t^{'} }^{t}\left[E_{0} +E_{1} \cos wt^{''} \right]dt^{''}  \right).
\end{eqnarray} 
Also,
\begin{equation}
	v_{z} (p)=\frac{\partial \varepsilon (p)}{\partial P_{z} } =\frac{\Delta _{z} d_{z} }{\hbar } \sin \frac{P_{z} d_{z} }{\hbar } \label{EQ_16},
\end{equation} 
and
\begin{eqnarray} 
\fl	v_{z} \left(p-e\int _{t-t^{'} }^{t}\left[E_{0} +E_{1} \cos wt^{''} \right]dt^{''}  \right) = \nonumber\\
	\frac{\Delta _{z} d_{z} }{\hbar } \left\{\sin \frac{P_{s} d_{s} }{\hbar } \cos \left(p-e\int _{t-t^{'} }^{t}\left[E_{0} +E_{1} \cos wt^{''} \right]dt^{''}  \right)\right\}\nonumber\\
-\cos \frac{P_{z} d_{z} }{\hbar } \sin \left(p-e\int _{t-t^{'} }^{t}\left[E_{0} +E_{1} \cos wt^{''} \right]dt^{''}  \right). \label{EQ_17} 
\end{eqnarray}

To calculate the carrier current density for a non-degenerate electron gas, the Boltzmann equilibrium distribution function \textit{f${}_{0}$(p)} is expressed as
\begin{equation}
f_{0} \left(p\right)=C\exp \left(\frac{\Delta _{s} \cos \frac{P_{s} d_{s} }{\hbar } +\Delta _{z} \cos \frac{P_{z} d_{z} }{\hbar } +\mu -\varepsilon _{0} }{kT} \right).  \label{EQ_18} 
\end{equation} 
Where C is found to be           
\begin{equation}
C=\frac{d_{s} d_{z} n_{0} }{2I_{0} \left(\Delta _{s}^{*} \right)I_{0} \left(\Delta _{z}^{*} \right)} \exp \left(-\frac{\mu -\varepsilon _{0} }{kT} \right)  \label{GrindEQ__19_} 
\end{equation} 
 and $n_{0}$ is the surface charge density, $I_{n}(x)$ is the modified Bessel function of order n defined by  $\Delta_s^{*} = \frac{\Delta_{s}}{kT}$ and $\Delta_{z}^{*} =\frac{\Delta _{z}^{} }{kT} $  and \textit{k} is Boltzmann's constant.\\

\noindent
Now, we substituted Eqs.(\ref{EQ_13}) and (\ref{EQ_18}) into Eqs.(\ref{EQ_11}) and (\ref{EQ_12}), and carried out the integrals. After cumbersome calculations the following expressions were obtain
\begin{eqnarray}
\fl	S' = -\sigma_{s} (E) \frac{1}{e}\Big\{(\varepsilon_0 - \mu)\sum_{n=-\infty}^{\infty}J_n^2(a) - \frac{\Delta_s}{2}\left(1 + 3\sum_{n=-\infty}^{\infty}J_n^2(a)\right)\left(\frac{I_0(\Delta^*_z)}{I_1(\Delta^*_z)} - \frac{2}{\Delta^*_s}\right)\nonumber\\
	-\Delta_s\frac{I_1(\Delta^*_z)}{I_0(\Delta^*_z)}\sum_{n=-\infty}^{\infty}J_n^2(a)\Big\}E^*_{sn}\nonumber\\
	- \sigma_{s} (E) \frac{k}{e^2}\Big\{\frac{(\varepsilon_0 - \mu)^2}{kT}\sum_{n=-\infty}^{\infty}J_n^2(a) - \frac{\Delta_s}{2}\frac{(\varepsilon_0 - \mu)}{kT}\left(1 + 3\sum_{n=-\infty}^{\infty}J_n^2(a)\right)\nonumber\\
	\times \left(\frac{I_0(\Delta^*_z)}{I_1(\Delta^*_z)} - \frac{2}{\Delta^*_s}\right) - 2\Delta_s\frac{(\varepsilon_0 - \mu)}{kT}\frac{I_1(\Delta^*_z)}{I_0(\Delta^*_z)}\sum_{n=-\infty}^{\infty}J_n^2(a)\nonumber\\
	+ \frac{\Delta_s\Delta^*_s}{2}\left(1 -  \frac{3I_0(\Delta^*_s)}{I_1(\Delta^*_s)} + \frac{6}{\Delta^*_{s^2}}\right)\left(1 + \sum_{n=-\infty}^{\infty}J_n^2(a)\right)\nonumber\\
	+ \frac{\Delta_s\Delta^*_s}{2}\left(\frac{I_0(\Delta^*_s)}{I_1(\Delta^*_s)} -  \frac{2}{\Delta^*_s}\right)\left(1 + 3\sum_{n=-\infty}^{\infty}J_n^2(a)\right) \nonumber\\
	+ \Delta_z\Delta^*_z\left(1 - \frac{I_1(\Delta^*_z)}{\Delta^*_zI_0(\Delta^*_z)}\right)\sum_{n = -\infty}^{\infty}J_n^2(a)\Big\}\nabla_sT \label{EQ_20}
\end{eqnarray}
\begin{eqnarray}
\fl	Z' = -\sigma_{z} (E) \frac{1}{e}\Big\{(\varepsilon_0 - \mu)\sum_{n=-\infty}^{\infty}J_n^2(a) - \frac{\Delta_z}{2}\left(1 + 3\sum_{n=-\infty}^{\infty}J_n^2(a)\right)\left(\frac{I_0(\Delta^*_z)}{I_1(\Delta^*_z)} - \frac{2}{\Delta^*_s}\right)\nonumber\\
	-\Delta_s\frac{I_1(\Delta^*_s}{I_0(\Delta^*_s)}\sum_{n=-\infty}^{\infty}J_n^2(a)\Big\}E^*_{zn}\nonumber\\
	- \sigma_{z} (E) \frac{k}{e^2}\Big\{\frac{(\varepsilon_0 - \mu)^2}{kT}\sum_{n=-\infty}^{\infty}J_n^2(a) - \frac{\Delta_z}{2}\frac{(\varepsilon_0 - \mu)}{kT}\left(1 + 3\sum_{n=-\infty}^{\infty}J_n^2(a)\right)\nonumber\\
	\times \left(\frac{I_0(\Delta^*_s)}{I_1(\Delta^*_s)} - \frac{2}{\Delta^*_z}\right) - 2\Delta_s\frac{(\varepsilon_0 - \mu)}{kT}\frac{I_1(\Delta^*_z)}{I_0(\Delta^*_z)}\sum_{n=-\infty}^{\infty}J_n^2(a)\nonumber\\
	+ \frac{\Delta_z\Delta^*_z}{2}\left(1 -  \frac{3I_0(\Delta^*_z)}{I_1(\Delta^*_z)} + \frac{6}{\Delta^*_{z^2}}\right)\left(1 + \sum_{n=-\infty}^{\infty}J_n^2(a)\right)\nonumber\\
	+ \frac{\Delta_z\Delta^*_s}{2}\left(\frac{I_0(\Delta^*_z)}{I_1(\Delta^*_z)} -  \frac{2}{\Delta^*_z}\right)\left(1 + 3\sum_{n=-\infty}^{\infty}J_n^2(a)\right) \nonumber\\
	+ \Delta_s\Delta^*_s\left(1 - \frac{I_1(\Delta^*_s)}{\Delta^*_sI_0(\Delta^*_s)}\right)\sum_{n = -\infty}^{\infty}J_n^2(a)\Big\}\nabla_zT	 \label{EQ_21}
\end{eqnarray}
Where we have defined $E^*_{sn}$ as 
\[
	E^*_{sn} = E_n + \nabla_s\frac{\mu}{e},  
\]
and $\sigma_i(E)$ as
\begin{equation}
	\sigma_s(E) = \frac{e^2\tau\Delta_sd_s^2n_0}{\hbar^2}\frac{I_1(\Delta^*_s)}{I_0(\Delta^*_s)},\qquad i = s,\,z
	\label{EQ_22}
\end{equation}
with $J_n(a)$ in Eqs.\ref{EQ_20} and \ref{EQ_21} is the bessel function of the $n$th order. $a = ed_sE_s/\omega \hbar$. $E_i$ is the amplitude of the laser. Substituting Eq.\ref{EQ_20} into Eq.\ref{EQ_9} gives circumferential thermal current density $q_c$ as
\begin{eqnarray}
\fl q_{c} = - \sigma_s(E)\frac{kT}{e} \sin \theta _h \cos \theta _h\nonumber\\
	\times \left\{\xi \sum _{n=-\infty }^{\infty }J_{n}^{2}(a) -\frac{\Delta _s^* }{2} B_s \left(1+3\sum _{n=-\infty }^{\infty }J_{n}^{2}  (a)\right) - \Delta^*_zA_z \sum _{n=-\infty }^{\infty }J_{n}^{2}  (a)\right\}E_{zn}^*\nonumber\\
	- \sigma_s(E)\frac{k^2T}{e^2} \sin \theta _h \cos \theta _h \left\{\xi^2 \sum_{n=-\infty }^{\infty }J_{n}^{2}(a) - \frac{\Delta _s^* }{2}\xi B_s \left(1+3\sum _{n=-\infty }^{\infty }J_{n}^{2}  (a)\right)\right. \nonumber\\
	- 2\Delta^*_z\xi A_z\sum_{n=-\infty}^{\infty}j_n^2(a) + \frac{\Delta^*_s)^2}{2}C_s\left(1 + \sum_{n=-\infty}^{\infty}j_n^2(a)\right) + \frac{\Delta^*_s\Delta^*_z}{2}B_sA_z\nonumber\\
	\times \left(1 + 3\sum_{n=-\infty}^{\infty}j_n^2(a)\right)\left. + (\Delta^*_s)^2\left(1 - \frac{A_z}{\Delta_z}\right)\sum_{n=-\infty}^{\infty}j_n^2(a) \right\}\Delta_{z}T. \label{EQ_23}
\end{eqnarray}
Also, substituting Eq.(\ref{EQ_21}) into Eq.(\ref{EQ_10}) gives axial thermal current density qz as
\begin{eqnarray}
\fl	q_{z} =-\frac{kT}{e} \left\{\sigma _{z} \left(E\right)\left[\xi \sum _{n=-\infty }^{\infty }J_{n}^{2}  \left(a\right)\right. \right. -\frac{\Delta _{z}^{*} }{2} B_{z} \left(1+3\sum _{n=-\infty }^{\infty }J_{n}^{2}  \left(a\right)\right)\left. -\Delta _{s}^{*} A_{s} \sum _{n=-\infty }^{\infty }J_{n}^{2}  \left(a\right)\right]\nonumber\\
	+\sigma _{s} \left(E\right)\sin ^{2} \theta _{h} \left[\xi \sum _{n=-\infty }^{\infty }J_{n}^{2}  \left(a\right)\right. -\frac{\Delta _{s}^{*} }{2} B_{s} \left(1+3\sum _{n=-\infty }^{\infty }J_{n}^{2}  \left(a\right)\right)\nonumber\\
	\left. \left. -\Delta _{z}^{*} A_{z} \sum _{n=-\infty }^{\infty }J_{n}^{2}  \left(a\right)\right]\right\}E_{zn}^{*} \nonumber\\
	- \frac{k^{2} T}{e^{2} } \left\{\sigma _{z} \left(E\right)\left[\xi ^{2} \sum _{n=-\infty }^{\infty }J_{n}^{2}  \left(a\right)\right. \right. - \frac{\Delta _{z}^{*} }{2} \xi B_{z} \left(1+3\sum _{n=-\infty }^{\infty }J_{n}^{2}(a)\right)\nonumber\\
	- 2\Delta _{s}^{*} \xi A_{s} \sum _{n=-\infty }^{\infty }J_{n}^{2}(a) + \frac{\left(\Delta _{z}^{*} \right)^{2} }{2} C_{z} \left(1 + \sum _{n=-\infty }^{\infty }J_{n}^{2}(a)\right) \nonumber\\
	+ \frac{\Delta^*_{z} \Delta^*_s}{2} A_{s} B_{z} \left(1+3\sum _{n=-\infty }^{\infty }J_{n}^{2}(a)\right) + (\Delta^*_s)^{2} \left(1 - \frac{A_{s} }{\Delta^*_s} \right)\left. \sum _{n=-\infty }^{\infty }J_{n}^{2}(a)\right]\nonumber\\
	+\sigma_{s} (E)\sin ^{2} \theta _{h} \left[\xi ^{2} \sum _{n=-\infty }^{\infty }J_{n}^{2}(a)\right. - \frac{\Delta^*_s }{2} \xi B_{s} \left(1+3\sum _{n=-\infty }^{\infty }J_{n}^{2}(a)\right)\nonumber\\
	- 2\Delta _{z}^{*} \xi A_{z} \sum _{n=-\infty }^{\infty }J_{n}^{2}(a) + \frac{\left(\Delta _{s}^{*} \right)^{2} }{2} C_{s} \left(1+\sum _{n=-\infty }^{\infty }J_{n}^{2}(a)\right) +  \frac{\Delta _{s} ^{*} \Delta _{z}^{*} }{2} A_{z} B_{s} \nonumber\\  
	\times \left(1+3\sum _{n=-\infty }^{\infty }J_{n}^{2}(a)\right) +\left(\Delta _{z}^{*} \right)^{2} \left(1-\frac{A_{z} }{\Delta _{z}^{*} } \right)\left. \left. \sum _{n=-\infty }^{\infty }J_{n}^{2}  \left(a\right)\right]\right\}\nabla _{z} T.  \label{EQ_24} 
\end{eqnarray}
Here we have used the following definitions
\begin{equation}
\fl	\xi =\frac{\varepsilon _{0} -\mu }{kT}, \quad A_i = \frac{I_{1}(\Delta^*_i)}{I_{0}(\Delta _{i}^{*}}, \quad B_{i} = \frac{I_{0}(\Delta^*_i)}{I_{1}(\Delta^*_i)} -\frac{2}{\Delta^*_i}, \quad C_i = 1 - \frac{3I_0(\Delta^*_i)}{\Delta^*_i I_1(\Delta^*_i )} + \frac{6}{\Delta^*_i{^2} }.       \label{EQ_25}
 \end{equation}

 \noindent
 The circumferential $\chi_{ec}$ and axial $\chi_{ez}$ components of the electron thermal conductivity in the CNT are obtained from Eqs (\ref{EQ_23}) and (\ref{EQ_24}) respectively. In fact the coefficients of the temperature gradient in these equations define $\chi_{ec}$ and $\chi_{ez}$ as follows,

\begin{eqnarray}        
\fl	\chi _{ec} =\sigma _{s} \left(E\right)\frac{k^{2} T}{e^{2} } \sin \theta _{h} \cos \theta _{h} \left\{\xi ^{2} \sum _{n=-\infty }^{\infty }J_{n}^{2}  \left(a\right)\right. -\frac{\Delta _{s}^{*} }{2} \xi B_{s} \left(1+3\sum _{n=-\infty }^{\infty }J_{n}^{2}  \left(a\right)\right)\nonumber\\     
	-2\Delta _{z}^{*} \xi A_{z} \sum _{n=-\infty }^{\infty }J_{n}^{2}  \left(a\right)+\frac{\left(\Delta _{s}^{*} \right)^{2} }{2} C_{s} \left(1+\sum _{n=-\infty }^{\infty }J_{n}^{2}  \left(a\right)\right)\nonumber\\
	+\frac{\Delta _{s} ^{*} \Delta _{z}^{*} }{2} B_{s} A_{z} \left(1+3\sum _{n=-\infty }^{\infty }J_{n}^{2}  \left(a\right)\right)+\left(\Delta _{z}^{*} \right)^{2} \left(1-\frac{A_{z} }{\Delta _{z}^{*} } \right)\left. \sum _{n=-\infty }^{\infty }J_{n}^{2}  \left(a\right)\right\}
 \end{eqnarray}
 
\begin{eqnarray} 
\fl	\chi _{ez} =\frac{k^{2} T}{e^{2} } \left\{\sigma _{z} \left(E\right)\left[\xi ^{2} \sum _{n=-\infty }^{\infty }J_{n}^{2}  \left(a\right)\right. \right. -\frac{\Delta _{z}^{*} }{2} \xi B_{z} \left(1+3\sum _{n=-\infty }^{\infty }J_{n}^{2}  \left(a\right)\right)-2\Delta _{s}^{*} \xi A_{s} \sum _{n=-\infty }^{\infty }J_{n}^{2}  \left(a\right)\nonumber\\
	+\frac{\left(\Delta _{z}^{*} \right)^{2} }{2} C_{z} \left(1+\sum _{n=-\infty }^{\infty }J_{n}^{2}  \left(a\right)\right)+\frac{\Delta _{z} ^{*} \Delta _{s}^{*} }{2} A_{s} B_{z} \left(1+3\sum _{n=-\infty }^{\infty }J_{n}^{2}  \left(a\right)\right)\nonumber\\
	+\left(\Delta _{s}^{*} \right)^{2} \left(1-\frac{A_{s} }{\Delta _{s}^{*} } \right)\left. \sum _{n=-\infty }^{\infty }J_{n}^{2}  \left(a\right)\right]+\sigma _{s} \left(E\right)\sin ^{2} \theta _{h} \left[\xi ^{2} \sum _{n=-\infty }^{\infty }J_{n}^{2}  \left(a\right)\right. \nonumber\\
	-\frac{\Delta _{s}^{*} }{2} \xi B_{s} \left(1+3\sum _{n=-\infty }^{\infty }J_{n}^{2}  \left(a\right)\right)-2\Delta _{z}^{*} \xi A_{z} \sum _{n=-\infty }^{\infty }J_{n}^{2}  \left(a\right)\nonumber\\
	+\frac{\left(\Delta _{s}^{*} \right)^{2} }{2} C_{s} \left(1+\sum _{n=-\infty }^{\infty }J_{n}^{2}  \left(a\right)\right)+\frac{\Delta _{s} ^{*} \Delta _{z}^{*} }{2} A_{z} B_{s} \left(1+3\sum _{n=-\infty }^{\infty }J_{n}^{2}  \left(a\right)\right)\nonumber\\
	+\left(\Delta _{z}^{*} \right)^{2} \left(1-\frac{A_{z} }{\Delta _{z}^{*} } \right)\left. \left. \sum _{n=-\infty }^{\infty }J_{n}^{2}  \left(a\right)\right]\right\}.  \label{EQ_27} 
  \end{eqnarray}

\noindent 
In summary, the analytical expressions obtained for the thermal current density \textit{q} and electron thermal conductivity $\chi $ depend on the geometric chiral angle $\theta_{h}$, temperature T, the real overlapping integrals for jumps along the tubular axis $\Delta_{z}$ and the base helix $\Delta_{s}$. 

\noindent 
When the Laser source is switched off i.e. E${}_{s}$ = 0\textit{,} the circumferential and axial components of the electron thermal conductivity expressions in Eqs.(\ref{EQ_26}) and (\ref{EQ_27}) reduces to
\begin{eqnarray}
\fl	\chi _{ec} =\sigma _{s} \left(E\right)\frac{k^{2} T}{e^{2} } \sin \theta _{h} \cos \theta _{h} \left\{\xi ^{2} \right. -2\Delta _{s}^{*} \xi B_{s}  -2\Delta _{z}^{*} \xi A_{z} +\left(\Delta _{s}^{*} \right)^{2} C_{s}\nonumber\\
	+2\Delta _{s} ^{*} \Delta _{z}^{*} B_{s} A_{z} +\left(\Delta _{z}^{*} \right)^{2} \left(1-\frac{A_{z} }{\Delta _{z}^{*} } \right). \label{EQ_28} 
\end{eqnarray} 

\begin{eqnarray}
\fl	\chi _{ez} =\frac{k^{2} T}{e^{2} } \left\{\sigma _{z} \left(E\right)\left[\mathop{\xi ^{2} }\limits_{}^{} \right. \right. -2\Delta _{z}^{*} \xi B_{z} -2\Delta _{s}^{*} \xi A_{s} +2\left(\Delta _{z}^{*} \right)^{2} C_{z} +2\Delta _{z} ^{*} \Delta _{s}^{*} A_{s} B_{z} \nonumber\\
	+ \left. \left(\Delta _{s}^{*} \right)^{2} \left(1-\frac{A_{s} }{\Delta _{s}^{*} } \right)\right]+\sigma _{s} \left(E\right)\sin ^{2} \theta _{h} \left[\mathop{\xi ^{2} }\limits_{}^{} \right. -2\Delta _{s}^{*} \xi B_{s} -2\Delta _{z}^{*} \xi A_{z} \nonumber\\
	+2\left(\Delta _{s}^{*} \right)^{2} C_{s} +2\Delta _{s} ^{*} \Delta _{z}^{*} A_{z} B_{s} +\left. \left. \left(\Delta _{z}^{*} \right)^{2} \left(1-\frac{A_{z} }{\Delta _{z}^{*} } \right)\right]\right\}.  \label{EQ_29}
\end{eqnarray} 
which are obtained in \cite{13}.

\section{Results, Discussion and Conclusion}
The Boltzmann transport equation was utilized to obtain expressions of the thermal current density and electron thermal conductivity of chiral SWNT.

We observed from \eref{EQ_26} and \eref{EQ_27} that the electron thermal conductivity of a chiral CNT is dependent on the electric field $E_{s}$, temperature T, GCA $\theta_{h}$, and the overlapping integrals $\Delta_{s}$ and $\Delta_{z}$ for jumps along the circumferential and axial directions.  Using MATLAB we sketched \eref{EQ_26} and \eref{EQ_27} to show how these parameters affect the electron thermal conductivity of a chiral CNT.

Figure (1a) represents the dependence of circumferential electron thermal conductivity, $\chi_{c}$, on temperature for a fixed value of $\Delta_{s}$ = 0.010eV and values of $\Delta_{z}$ varied from 0.010eV to 0.014eV.  We noticed that the relationship between $\chi_{c}$ and $T$ is nonlinear and indicates a positive slope at low temperatures and negative slope at high temperatures. The physical interpretation to the part of the graph showing positive slope is that more electrons are thermally generated to transport heat through the chiral CNT. The peak of the graph indicates the threshold temperature at which electron and heat transport through the chiral CNT is maximum. The negative slope of the graph shows that as temperature exceeds the threshold value, carbon atoms are energized to vibrate faster thereby scattering the electrons carrying thermal energies through the chiral CNT. At temperatures above 300K, $\chi_{c}$ assumes a lower constant value for all values of $\Delta_{z}$. The peak values of $\chi_{c}$ were found to decrease as  $\Delta_{z}$ increases. Figure (1b) shows the dependence of $\chi_{c}$, on temperature for GCA $\theta_{h}$ varied between 1.2$^{o}$ and 2.0$^{o}$. It was noted that the values of $\chi_{c}$ increases with increasing GCA $\theta_{h}$. 

Figure (1c) represents the dependence of axial electron thermal conductivity, $\chi_{z}$, on temperature T for a fixed value of $\Delta_{s}$ = 0.010eV and values of $\Delta_{z}$ varied from 0.010eV to 0.015eV. Like $\chi_{c}$, the relationship between $\chi_{z}$ and T is also found to be nonlinear and indicates a positive slope at low temperatures and negative slope at high temperatures. It was observed that as $\Delta_{z}$ increases, the values of $\chi_{z}$ also increase. It is quite interesting to note that the values of $\chi_{z}$ are much larger as compared with those of $\chi_{c}$. 

Electron thermal conductivities $\chi_{c}$ and $\chi_{z}$ dependence on temperature in the presence and also absence of Laser is sketched and presented as Figures (2a) and (2b). In comparison, we noted that the presence of laser causes a drastic reduction in $\chi_{c}$ and $\chi_{z}$. The reason is that the Laser source $E_{s}$ energizes the carbon atoms within the walls of the CNT and set them vibrating at large amplitudes which tend to scatter the electrons carrying thermal energy.

Figures (3a) and (3b) illustrate the behaviour of $\chi_{c}$ and $\chi_{z}$ as the Laser source $E_{s}$ is varied. We noticed that as the Laser source increases $\chi_{c}$ and $\chi_{z}$ drops off sharply and oscillates towards larger $E_{s}$ values. As $E_{s}$ values become larger, the amplitudes of oscillation decrease.
\begin{figure}[htb!]
	\centerline{\includegraphics[width=11in,height=15in]{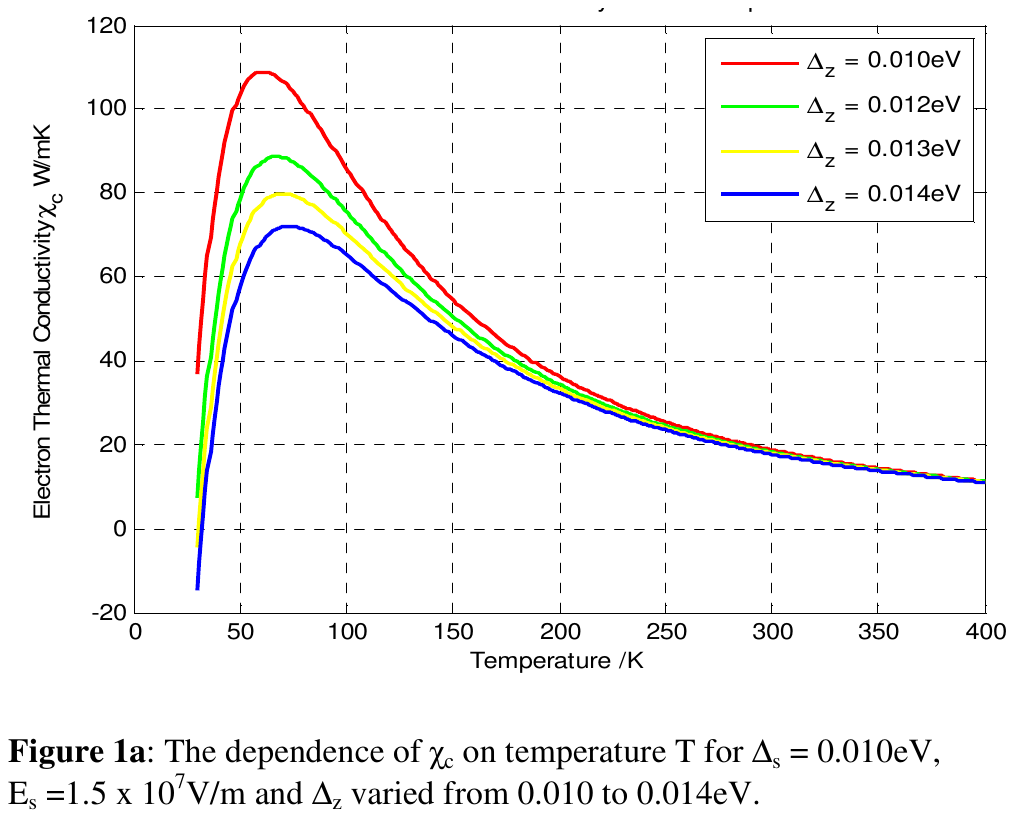}}
\end{figure}

\begin{figure}[htb!]
	\centerline{\includegraphics[width=11in,height=15in]{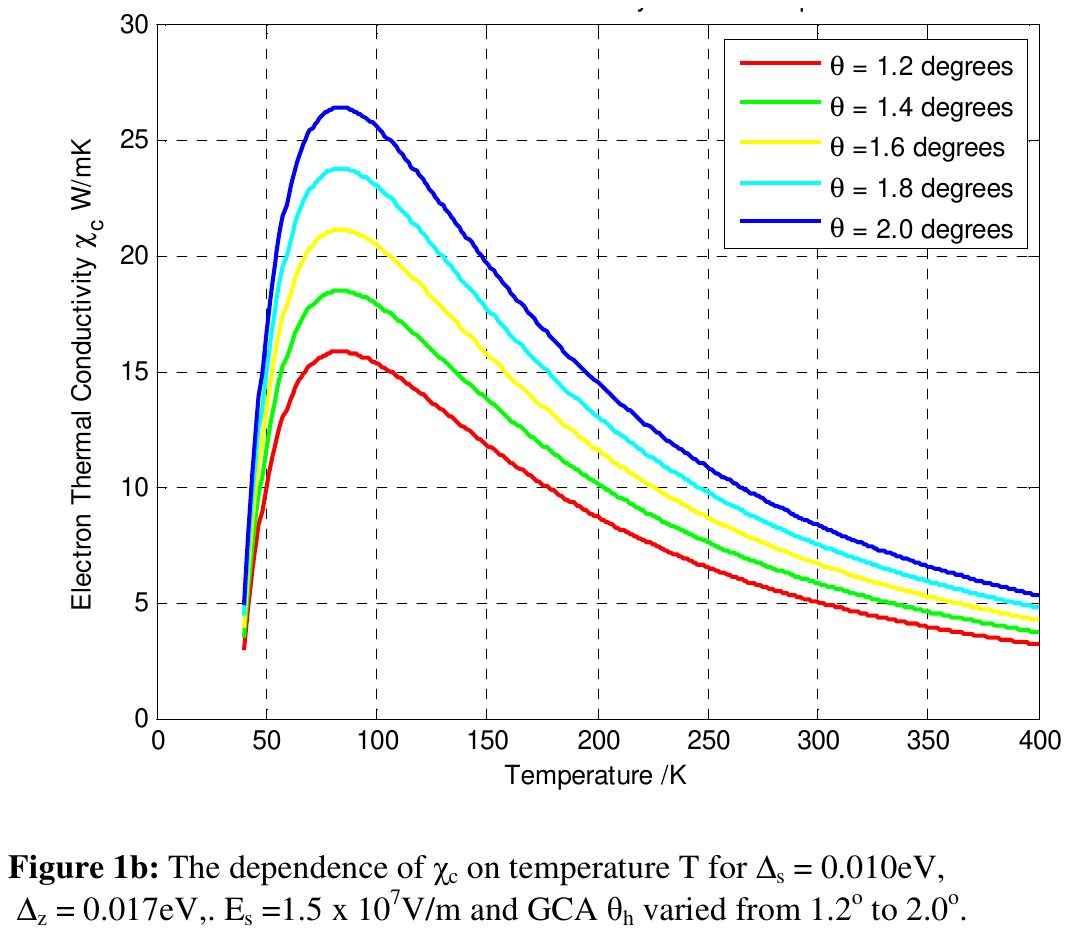}}
\end{figure}

\begin{figure}[htb!]
	\centerline{\includegraphics[width=11in,height=15in]{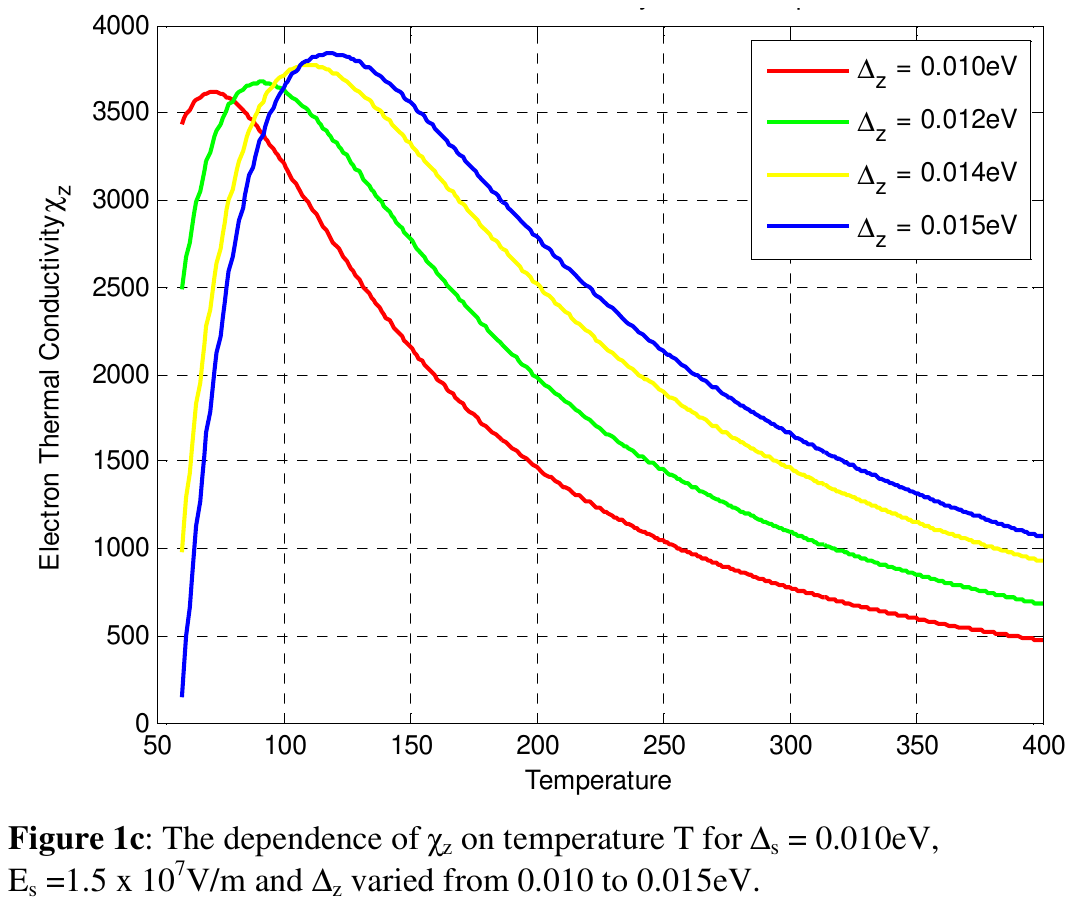}}
\end{figure}

\begin{figure}[htb!]
	\centerline{\includegraphics[width= 11in,height = 15in]{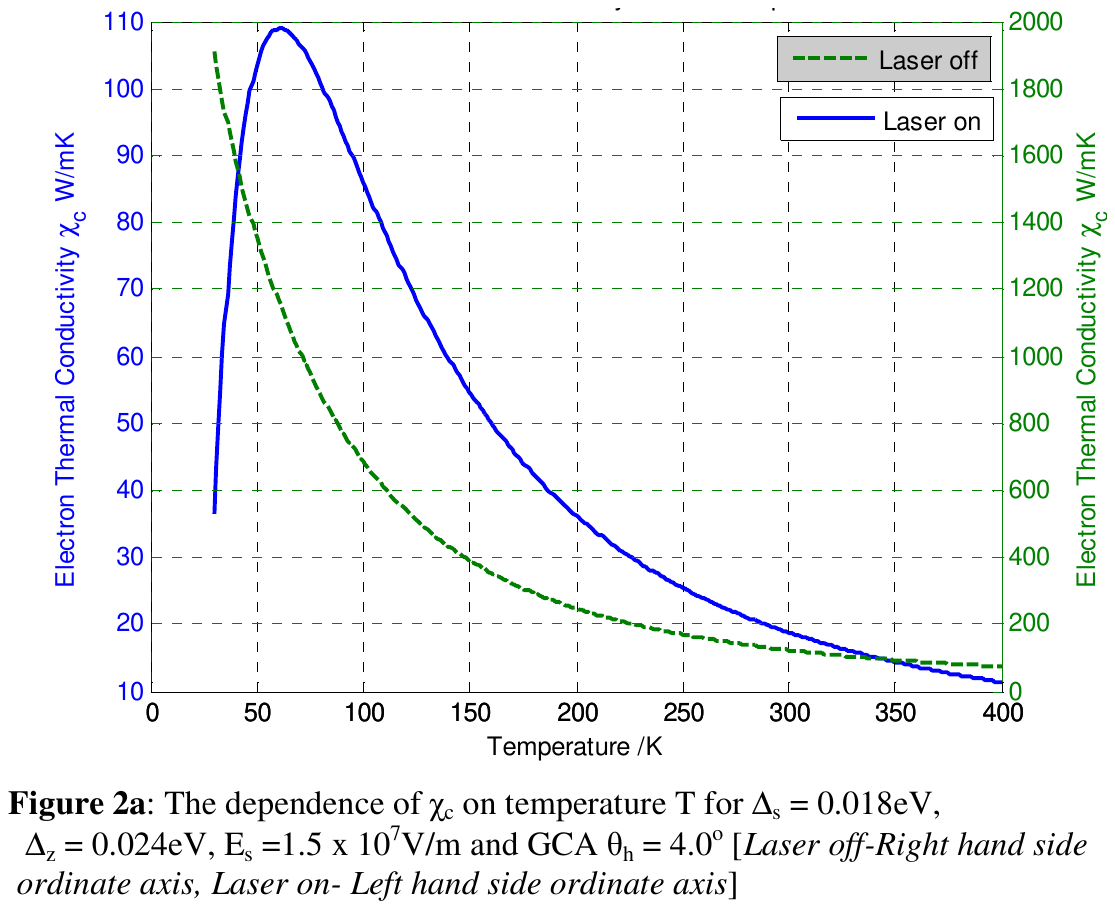}}
\end{figure}

\begin{figure}[htb!]
	\centerline{\includegraphics[width = 11.0in, height = 15in]{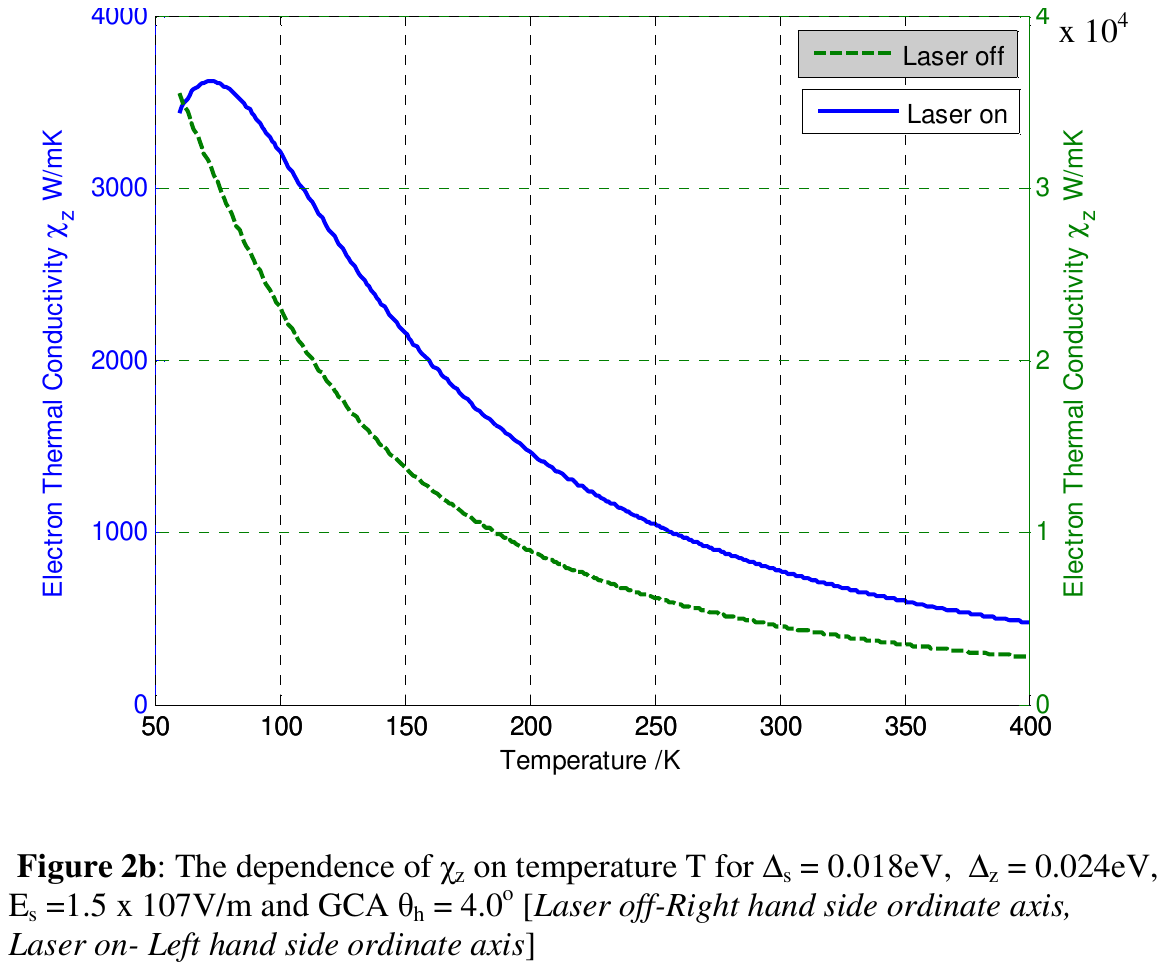}}
\end{figure}

\begin{figure}[htb!]
	\centerline{\includegraphics[width=11in,height=15in]{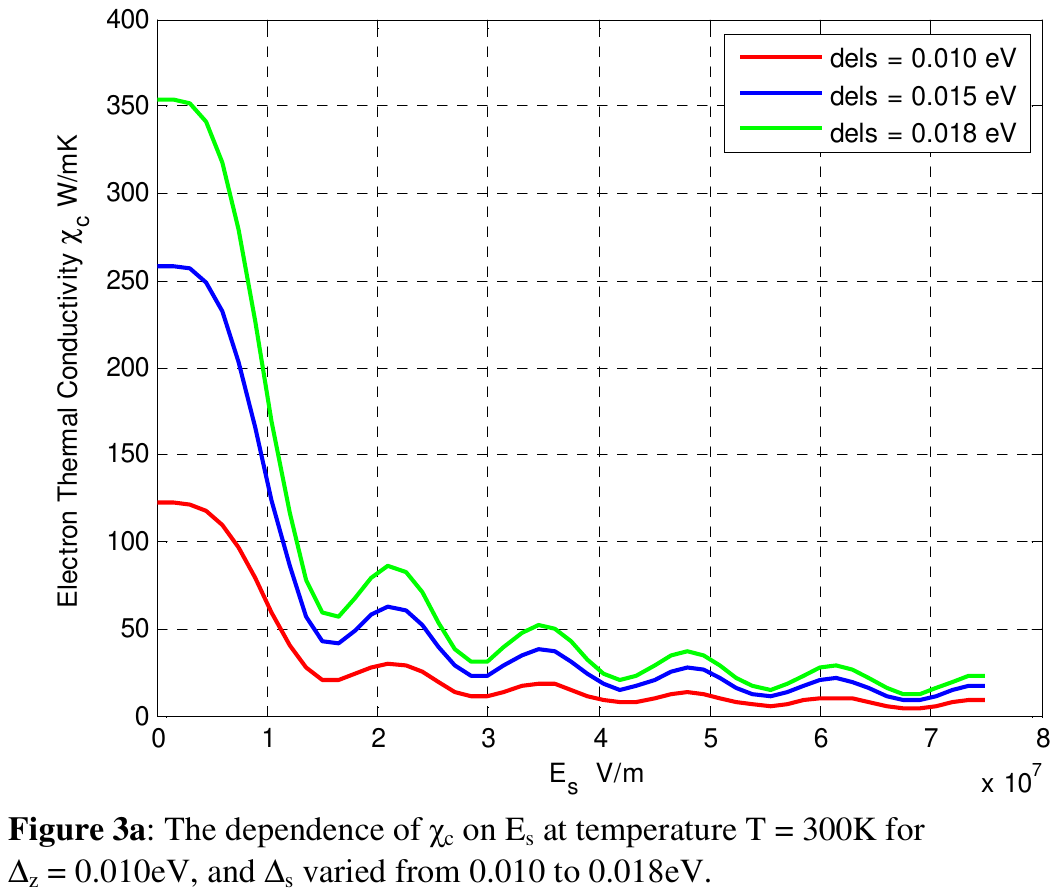}}
\end{figure}

\begin{figure}[htb!]
	\centerline{\includegraphics[width=11in,height=15in]{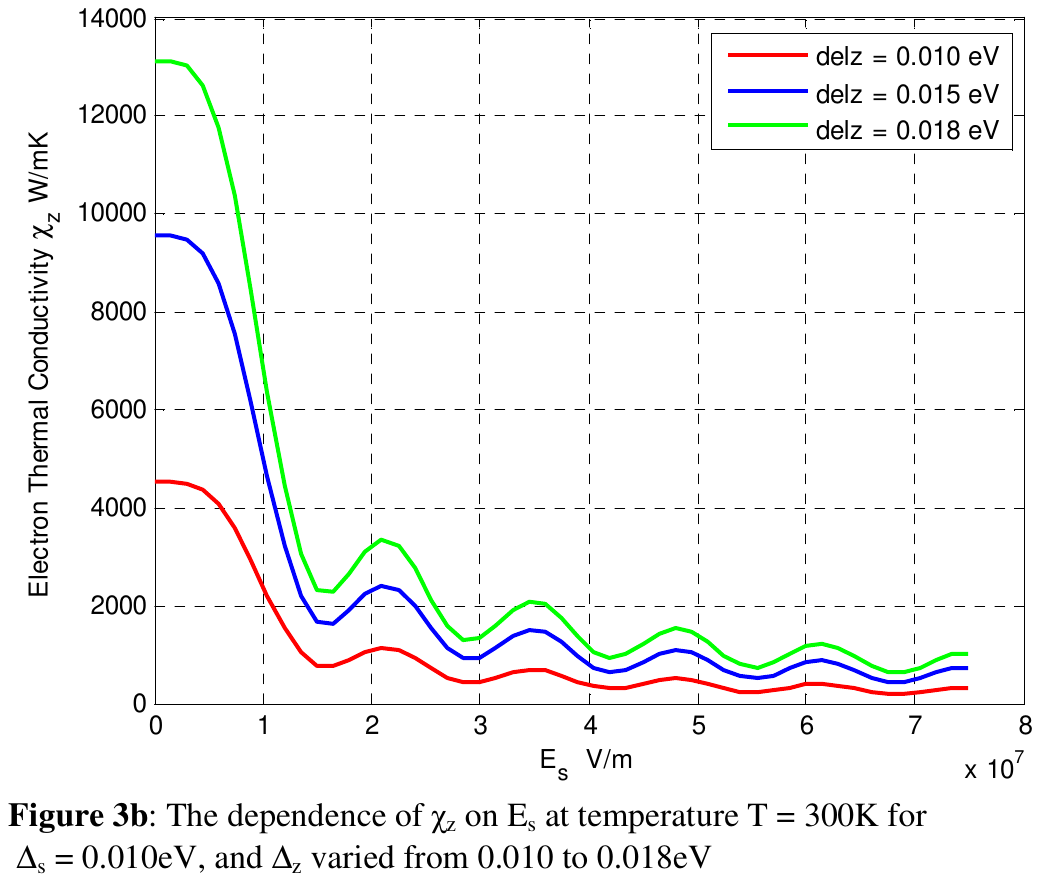}}
\end{figure}

\section{Conclusions}
The electron thermal conductivity $\chi$ of chiral CNT induced with monochromatic laser have been investigated. The chiral CNT parameters $\Delta_{s}$ $\Delta_z$, $\theta_{h}$, and the laser source $E_{s}$ were found to have influence on the electron thermal conductivity $\chi$ of chiral CNT.

 Our results show that a greater percentage of the electron and heat transport is along the axis of the chiral CNT.  It was observed that an increase in $\Delta_{z}$ causes $\chi_{c}$ to decrease and $\chi $${}_{z}$ to increase. Also an increase in $\theta $${}_{h}$ made $\chi $${}_{c}$ to rise but had no effect on $\chi $${}_{z}$. The parameters $\chi $${}_{c}$ and $\chi $${}_{z}$ were also found to be oscillating when the laser source E${}_{s}$ was varied. The results obtained indicated that the laser source caused a drastic reduction in the $\chi $ values. The reduced values recorded for $\chi $${}_{c}$ and $\chi $${}_{z}$ is a clear indication that the laser retains heat at the junctions of the chiral CNT which helps to maintain a large temperature gradient.

\newpage



\begin{thebibliography}{10}
\bibitem{1} 
Kaye G.W.C., and Laby T.H., Tables of Physical and Chemical Constant, sixteenth ed., \textit{Longman}, London, 1995.

\bibitem{2} 
Iijima S., \textit{Nature} 56 354 (1991). 

\bibitem{3}  
Hone J., Whitney M., Piskoti C., and Zettl A \textit{Phys, Rev}. B \textbf{59} 2514 (1999). 

\bibitem{4} 
Bonard J.M., Salvetat J.P, Stockli T.,and De Heer W.A., \textit{Appl. Phys. Lett.} 73 918 (1998). 

\bibitem{5} 
Wang Q.H., Corrigan T.D., Dai T.Y., and Chang R.P.H., \textit{Appl. Phys. Lett}. \textbf{70 }3308 (1997).

\bibitem{6} 
De Heer W.A, Chatelain A.,and Ugarte D., \textit{Science} \textbf{270} 1179 (1995)

\bibitem{7} 
Rinzler A.G., Hafner J.H., Nikolaev P., Lou L., Kim S.G., Tomanekb D., Nordlander P., Colbert D.T., and Smalley R.E., \textit{Science} \textbf{269 }1550. (1995) 

\bibitem{8} 
Semet V., Thien Binh Vu, Vincent P., Guillot D., Teo K.B.K, Chhowalla M., Amaratunga G.A.J., andMilne W.I., \textit{Appl. Phys. Lett}. \textbf{81} 343 (2002)

\bibitem{9} 
Dillon A.C., Jones K.M., Bekkedahl T.A, Kiag C.H., Bethune D.S., and Heben M.J., \textit{Nature} \textbf{386 }377 (1997).

\bibitem{10}
Chico L., Crespi V.H., Benedict L.X. , Louie S.G., and Cohen M.L., \textit{Phys. Rev. Lett}. \textbf{76 }971 (1996).

\bibitem{11}
Derycke V., Martel R., Appenzeller and J., Avouris Ph., \textit{Nano Lett}, \textbf{9 }453\textbf{ }(2001).

\bibitem{12}
Berber S., Kwon Y.K., and Tomanek D., \textit{Phys. Rev. Lett}. \textbf{84 }4613\textbf{ }(2000). 

\bibitem{13}
 Mensah S.Y, Allotey F.K.A, Nkrumah G.,and Mensah N.G., \textit{Physical E} \textbf{23} 152 (2004).

\bibitem{14} 
Mensah N. G., Nkrumah G., Mensah S. Y., Allotey F. K. A. Temperature Dependence of the Thermal Conductivity in Chiral Carbon Nanotubes, \textit{Phys Lett A }\textbf{329 }369 (2004).

\bibitem{15} 
S.Y.Mensah, A. Twum, N. G. Mensah, K. A. Dompreh, S. S. Abukari, G. Nkrumah-Buandoh,  Effect of laser on thermopower of chiral carbon nanotube, Mesoscale and Nanoscale Physics (cond-mat.mes-hall) arXiv:1104.1913v1 (2011)      

\bibitem{16}
Slepyan G. Ya., Maksimenko S. A., Lakhtakia A., Yevtushenko O. M., and Gusakov A. V., \textit{Phys. Rev}. B\textit{ }\textbf{57} 16 9485 (1998).

\bibitem{17} 
Yevtushenko O. M., Ya Slepyan G., Maksimenko S. A, Lakhtakia A. and Romanov D. A., \textit{Phys. Rev. Lett}. \textbf{79}, 1102 (1997)

\bibitem{18} 
Miyamoto Y., Louie S. G. and Cohen M. L., \textit{Phys. Rev. Lett}\textbf{.\textit{ }76} 2121 (1996).

\bibitem{19} 
Mensah S.Y., Allotey F.K.A., Mensah N.G. and Nkrumah G., \textit{J. Phys}. \textbf{13 }5653 (2001)\textit{}

\bibitem{20} 
Romanov D. A. and Kibis O. V, \textit{Phys. Lett. }\textbf{A 178}, 335 (1993)\textit{}

\bibitem{21} 
Yi-Chun Su and Wen-Kuang Hsu, \textit{Appl. Phys. Lett}. \textbf{87}, 233112 (2005)\textit{}
\end{thebibliography}
\end{document}